\documentclass[twoside,twocolumn,9pt]{article}
\usepackage{extsizes}
\usepackage[super,sort&compress,comma]{natbib} 
\usepackage[version=3]{mhchem}
\usepackage[left=1.5cm, right=1.5cm, top=1.785cm, bottom=2.0cm]{geometry}
\usepackage{balance}
\usepackage{times,mathptmx}
\usepackage{sectsty}
\usepackage{graphicx} 
\usepackage{lastpage}
\usepackage[format=plain,justification=justified,singlelinecheck=false,font={stretch=1.125,small,sf},labelfont=bf,labelsep=space]{caption}
\usepackage{float}
\usepackage{fancyhdr}
\usepackage{fnpos}
\usepackage[english]{babel}
\addto{\captionsenglish}{%
  \renewcommand{\refname}{Notes and references}
}
\usepackage{array}
\usepackage{droidsans}
\usepackage{charter}
\usepackage[T1]{fontenc}
\usepackage[usenames,dvipsnames]{xcolor}
\usepackage{setspace}
\usepackage[compact]{titlesec}
\usepackage{hyperref}

\usepackage{xcolor}

\usepackage{epstopdf}

\definecolor{cream}{RGB}{222,217,201}

\begin{document}

\pagestyle{fancy}
\thispagestyle{plain}
\fancypagestyle{plain}{

\fancyhead[C]{\includegraphics[width=18.5cm]{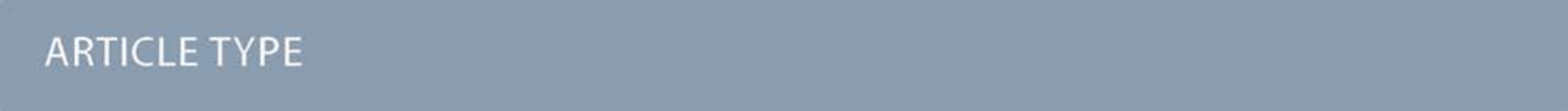}}
\fancyhead[L]{\hspace{0cm}\vspace{1.5cm}\includegraphics[height=30pt]{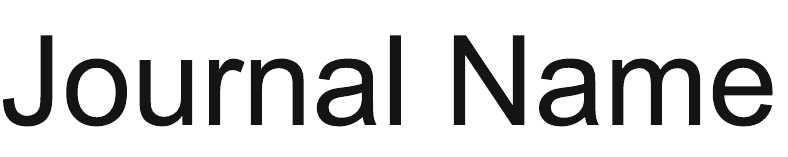}}
\fancyhead[R]{\hspace{0cm}\vspace{1.7cm}\includegraphics[height=55pt]{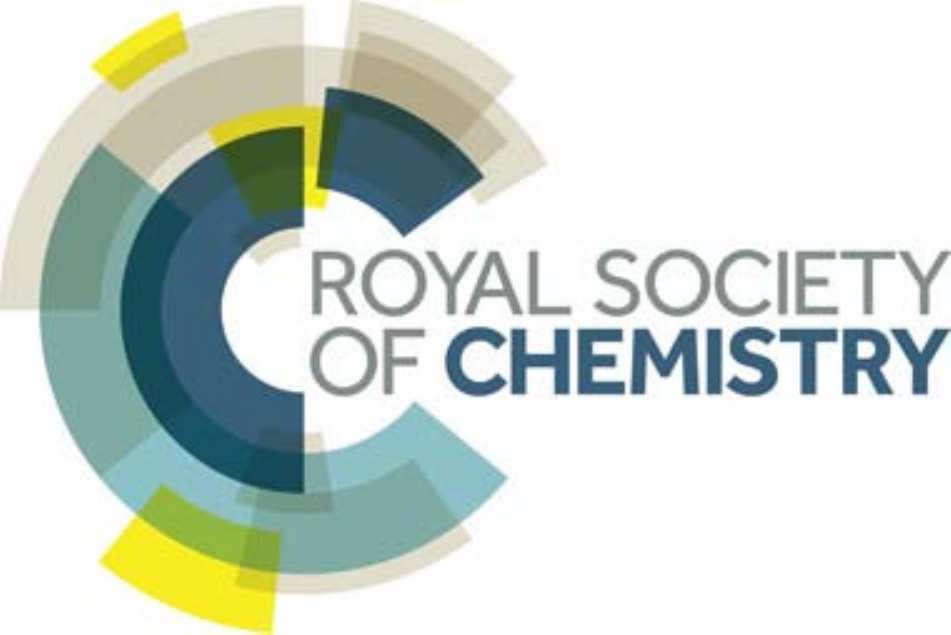}}
\renewcommand{\headrulewidth}{0pt}
}

\makeFNbottom
\makeatletter
\renewcommand\LARGE{\@setfontsize\LARGE{15pt}{17}}
\renewcommand\Large{\@setfontsize\Large{12pt}{14}}
\renewcommand\large{\@setfontsize\large{10pt}{12}}
\renewcommand\footnotesize{\@setfontsize\footnotesize{7pt}{10}}
\makeatother

\renewcommand{\thefootnote}{\fnsymbol{footnote}}
\renewcommand\footnoterule{\vspace*{1pt}%
\color{cream}\hrule width 3.5in height 0.4pt \color{black}\vspace*{5pt}} 
\setcounter{secnumdepth}{5}

\makeatletter 
\renewcommand\@biblabel[1]{#1}            
\renewcommand\@makefntext[1]%
{\noindent\makebox[0pt][r]{\@thefnmark\,}#1}
\makeatother 
\renewcommand{\figurename}{\small{Fig.}~}
\sectionfont{\sffamily\Large}
\subsectionfont{\normalsize}
\subsubsectionfont{\bf}
\setstretch{1.125} 
\setlength{\skip\footins}{0.8cm}
\setlength{\footnotesep}{0.25cm}
\setlength{\jot}{10pt}
\titlespacing*{\section}{0pt}{4pt}{4pt}
\titlespacing*{\subsection}{0pt}{15pt}{1pt}

\fancyfoot{}
\fancyfoot[LO,RE]{\vspace{-7.1pt}\includegraphics[height=9pt]{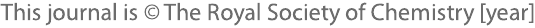}}
\fancyfoot[CO]{\vspace{-7.1pt}\hspace{13.2cm}\includegraphics{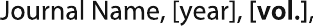}}
\fancyfoot[CE]{\vspace{-7.2pt}\hspace{-14.2cm}\includegraphics{head_foot/RF}}
\fancyfoot[RO]{\footnotesize{\sffamily{1--\pageref{LastPage} ~\textbar  \hspace{2pt}\thepage}}}
\fancyfoot[LE]{\footnotesize{\sffamily{\thepage~\textbar\hspace{3.45cm} 1--\pageref{LastPage}}}}
\fancyhead{}
\renewcommand{\headrulewidth}{0pt} 
\renewcommand{\footrulewidth}{0pt}
\setlength{\arrayrulewidth}{1pt}
\setlength{\columnsep}{6.5mm}
\setlength\bibsep{1pt}

\makeatletter 
\newlength{\figrulesep} 
\setlength{\figrulesep}{0.5\textfloatsep} 

\newcommand{\topfigrule}{\vspace*{-1pt}%
\noindent{\color{cream}\rule[-\figrulesep]{\columnwidth}{1.5pt}} }

\newcommand{\botfigrule}{\vspace*{-2pt}%
\noindent{\color{cream}\rule[\figrulesep]{\columnwidth}{1.5pt}} }

\newcommand{\dblfigrule}{\vspace*{-1pt}%
\noindent{\color{cream}\rule[-\figrulesep]{\textwidth}{1.5pt}} }

\makeatother

\twocolumn[
  \begin{@twocolumnfalse}
\vspace{3cm}
\sffamily
\begin{tabular}{m{4.5cm} p{13.5cm} }

\includegraphics{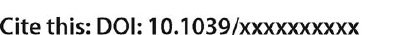} & 
\noindent\LARGE{\textbf{Transition Path Times in Asymmetric Barriers}} \\
\vspace{0.3cm} & \vspace{0.3cm} \\

 & \noindent\large{Michele Caraglio,\textit{$^{a,b}$} Takahiro
 Sakaue,\textit{$^{c,d}$} and Enrico Carlon\textit{$^{a\ddag}$}} \\

\includegraphics{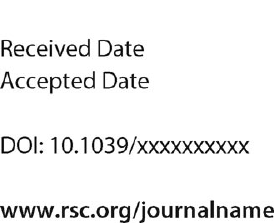} & \noindent\normalsize{
Biomolecular conformational transitions are usually modeled as barrier
crossings in a free energy landscape. The transition paths connect
two local free energy minima and transition path times (TPT) are the
actual durations of the crossing events. The simplest model employed to
analyze TPT and to fit empirical data is that of a stochastic particle
crossing a parabolic barrier. Motivated by some disagreement between
the value of the barrier height obtained from the TPT distributions as
compared to the value obtained from kinetic and thermodynamic analyses,
we investigate here TPT for barriers which deviate from the symmetric
parabolic shape. We introduce a continuous set of potentials, that
starting from a parabolic shape, can be made increasingly asymmetric
by tuning a single parameter. The TPT distributions obtained in the
asymmetric case are very well-fitted by distributions generated by
parabolic barriers. The fits, however, provide values for the barrier
heights and diffusion coefficients which deviate from the original input
values. We show how these findings can be understood from the analysis of
the eigenvalues spectrum of the Fokker-Planck equation and highlight
connections with experimental results.} \\

\end{tabular}

 \end{@twocolumnfalse} \vspace{0.6cm}
 
]

\renewcommand*\rmdefault{bch}\normalfont\upshape
\rmfamily
\section*{}
\vspace{-1cm}


\footnotetext{\textit{$^{a}$~KU Leuven, Soft Matter and Biophysics Unit, 
Celestijnenlaan 200D, B-3001 Leuven, Belgium.}}
\footnotetext{\textit{$^{b}$~Institut f\"ur Theoretische Physik,
Universit\"at Innsbruck, Technikerstra{\ss}e  21A, A-6020 Innsbruck,
Austria.}} 
\footnotetext{\textit{$^{c}$~Department of Physics and Mathematics,
Aoyama Gakuin University, 5-10-1 Fuchinobe, Chuo-ku, Sagamihara, Kanagawa
252-5258, Japan.}}
\footnotetext{\textit{$^{d}$~PRESTO, Japan Science and Technology Agency,
4-1-8 Honcho Kawaguchi, Saitama 332-0012, Japan.}}


\footnotetext{\ddag~Corresponding author: enrico.carlon@kuleuven.be}



\section{Introduction}

Conformational transitions in complex molecular systems are typically
described by the dynamics of a reaction coordinate in a free energy
landscape. Two local free energy minima correspond to two distinct
states of the system, say A and B. The transitions between the two
occur through the crossing of a free energy barrier. In this process
two different time scales can be identified. The first is the dwell
time, which is the average time the system stays in A (B) before making
a transition to B (A). A second time scale is the transition path time
(TPT) which is the time needed to actually cross the barrier from A
to B. The system spends most of the time fluctuating around the local
minima A or B, while transitions between them are very rapid. The average
dwell times can therefore be considerably longer that the average TPT,
as indeed seen in experiments of protein and nucleic acids folding
\cite{chun12,neup12}.

Considerable attention has been devoted in the past decade to study the
properties of TPT, both in experiments \cite{chun09,true15, neup16,
hoff19, glad19} and in theories \cite{zhan07, chau10, orla11, kim15,
poll16, bere17, lale17, cara18, bere19}, see also Ref. \cite{hoff19b}
for a recent review. The analysis of TPT has attracted quite some interest
in the past as it is expected to shed some light on the strong molecular
rearrangements occuring close to the top of the free energy barrier. A
central model is that of a diffusive particle crossing a parabolic
potential barrier $V(x)=-kx^2/2$, where $x$ denotes a reaction
coordinate and $k>0$ (note that also paths crossings a linear symmetric
cusped potential barrier have been recently considered \cite{bere17,
bere19}). Given a generic potential barrier $V(x)$ one defines
transition paths as those originating at one side of the barrier in a
position $x=x_1$ and end at the opposite side at $x=x_2$. Absorbing
boundary conditions are imposed at $x=x_1$ and $x=x_2$ to eliminate
the recrossing events at the begin and end points of the trajectories.
In practice, however, free boundary conditions are used \cite{zhan07,
lale17}, as these are analytically simpler to implement and converge to
the absorbing boundaries results in the high barrier limit ($\Delta U \gg
k_B T$, with $\Delta U$ the barrier height, $T$ the temperature and $k_B$
the Boltzmann constant) where recrossings are extremely rare.

For parabolic barriers exact TPT distributions were derived in the free
boundary conditions case \cite{zhan07}, providing simple expressions which
have been used to fit experimental data \cite{neup16}. Calculations with
parabolic barriers were also extended to the case of inertia \cite{lale17}
and in the presence of memory or active forces \cite{poll16, sati17,
carl18}. Absorbing boundary conditions were recently implemented in the
parabolic model as well \cite{cara18}. Although any smooth potential
can be approximated by a parabola sufficiently close to its maximum, this
approximation may hold only for a very small reaction coordinate range. It
is therefore interesting to analyze some wider set of potentials departing
from a perfect parabolic shape, which is the aim of the present paper. We
focus here on a barrier which is composed of a parabolic part joined to
a linear part. In the model the degree of asymmetry of the potential
can be tuned starting from the fully symmetric parabolic shape to a
strongly asymmetric linear barrier. TPT distributions are obtained by
mapping the problem into one dimensional effective Schr\"odinger equation
\cite{risken}. As both parabolic and linear potentials are solvable, one
just needs to fix the matching conditions at the boundary between them,
as done in quantum mechanical problems with discontinuous potentials.
We show how TPT distributions from asymmetric potential can always be
fitted by distributions obtained from parabolic potentials, showing that
the shape of the distribution is quite universal, i.e., insensitive to
the actual potential shape. The fit however leads to estimates of the
barrier height and of the diffusion coefficient that deviate from the
original input values.  We discuss the origin of these effects and their
relevance in view of contradicting estimates of barrier heights obtained
in recent DNA folding experiments.

\begin{figure}[t]
\centering
\includegraphics[width=0.40\textwidth]{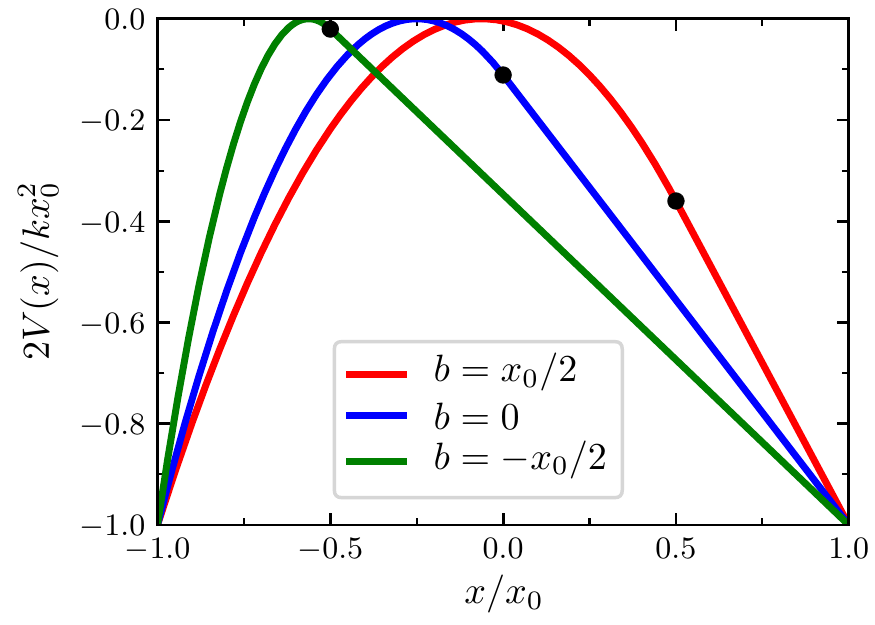}
\caption{Plot of the rescaled potential $2V(x)/kx_0^2$ vs. rescaled
distance $x/x_0$ defined by Eq.~\eqref{def:Vx} for $b=x_0/2$, $b=0$
and $b=-x_0/2$. The symmetric parabolic potential corresponds to $b=x_0$
and, as the parameter $b$ decreases, the potential becomes increasingly
asymmetric. $V(x)$ is parabolic for $-x_0 \leq x \leq b$ and linear for
$b \leq x \leq x_0$.}
\label{fig:V}
\end{figure}

\section{The model}

We consider a particle diffusing in a potential $V(x)$ defined in the
interval $-x_0 \leq x \leq x_0$ and parametrized as follows
\begin{equation}
V(x) = - \frac{k}{2} 
\left(\frac{x_0}{x_0 + \widetilde{x}}\right)^2 f(x)
\label{def:Vx}
\end{equation}
with
\begin{eqnarray}
f(x) = \left\{
\begin{array}{lr}
(x-\widetilde{x})^2 &-x_0 \leq x \leq b\\
(2x-b-\widetilde{x})(b-\widetilde{x})   & b \leq x \leq x_0
\end{array}
\right.&&
\label{def:V}
\end{eqnarray}
Here
\begin{equation}
\widetilde{x} = -\frac{(x_0-b)^2}{4x_0} \leq 0
\label{def:xtilde}
\end{equation}
is the location of the maximum of the potential
$V(\widetilde{x})=0$. Figure~\ref{fig:V} shows a plot of $V(x)$ for
three different values of $b$. In the case $b=x_0$ the potential is
parabolic. As $b$ decreases, $V(x)$ becomes more asymmetric with an
increasing negative curvature $\kappa_p=-kx_0^2/(x_0+\widetilde{x})^2$
in the parabolic region and a decrease of slope $s_{\ell}=\kappa_p
(b-\widetilde{x})$ in the linear region (see Fig.~\ref{fig:V}). The
barrier height is independent of $b$ and given by $\Delta U = \frac{1}{2}
k x_0^2$. Note that, by construction, $V(x)$ and $V'(x)$ are continuous
in $x=b$.

The Fokker-Planck equation, describing the evolution of the distribution
function $P(x,t)$, is given by:
\begin{equation}
\partial_t P(x,t) = D \partial_x 
\left[ e^{-\beta V(x)} \partial_x
\left( e^{\beta V(x)} P(x,t) \right) 
\right] \equiv L_\text{FP} \, P(x,t)
\label{eq:FP}
\end{equation}
where $D$ is the diffusion coefficient $\beta = 1/k_BT$ and $L_\text{FP}$
defines the Fokker-Planck operator. We impose absorbing boundary
conditions, hence $P(\pm x_0,t)=0$ at all times.

We use separation of variables and seek solutions of \eqref{eq:FP}
of the form
\begin{equation}
P(x,t) = e^{-\lambda D t} \, e^{-\beta V(x)/2} \, q(x), 
\label{sep_var}
\end{equation}
The parameter $\lambda$ determines the decay rate of a solution and from
general properties of the Fokker-Planck equation one has $\lambda > 0$,
see \cite{risken}. Substituting \eqref{sep_var} in \eqref{eq:FP} we get
\begin{equation}
q''(x) - V_\text{eff} (x) \, q(x) = -\lambda \, q(x)
\label{eq:FP2}
\end{equation}
where 
\begin{eqnarray}
V_\text{eff} (x) = -\frac{\beta}{2} V''(x) + \frac{\beta^2}{4} 
\left[V'(x) \right]^2 
\label{eff}
\end{eqnarray}
defines an effective potential \cite{risken}. 
We note that \eqref{eq:FP2}
is formally identical to the Schr\"odinger equation for a quantum particle
in a potential $V_\text{eff}(x)$ (we have set the units $\hbar^2/2m =1$,
the potential thus defined has the dimension of a squared inverse length)
and energy $\lambda$. The advantage of the transformation \eqref{sep_var}
is that \eqref{eff} defines an eigenvalue problem with an hermitian
operator (note that $L_\text{FP}$ in \eqref{eq:FP} is not hermitian),
hence eigenfunctions form a orthogonal complete set:
\begin{equation}
\int_{-x_0}^{+x_0} q_n(x) q_m(x) \, dx = \delta_{nm}
\label{eq:ortho}
\end{equation}
with $n$ and $m$ ``quantum" numbers.
Plugging \eqref{def:Vx} into \eqref{eff} we obtain
\begin{eqnarray}
V_\text{eff} (x)
= \left\{
\begin{array}{cr}
\frac{\widetilde{k}}{2} +
\frac{\widetilde{k}^2}{4} (x-\widetilde{x})^2 
& -x_0 \leq x \leq b\\
\frac{\widetilde{k}^2}{4} \left(b-\widetilde{x}\right)^2 & 
b \leq x \leq x_0
\end{array}
\right. &&
\label{def:Veff}
\end{eqnarray}
where we have defined
\begin{equation}
\widetilde{k} = \beta k \left(\frac{x_0}{x_0 + \widetilde{x}}\right)^2
= \frac{16 \beta k x_0^4}{(x_0+b)^2 (3x_0-b)^2}
\label{def:ktilde}
\end{equation}
(using \eqref{def:xtilde}).  Note that $\widetilde{k} = -\beta
\kappa_{p}$, hence the curvature of $V_\text{eff}$ at $x \in [-x_0,b]$
is $\widetilde{k}^2/2 = (\beta \kappa_p)^2/2$, and the constant value
in $V_\text{eff}$ at $x \in [b, x_0]$ (bottom of \eqref{def:Veff}) is
written as $(\beta s_{\ell}/2)^2$ using the curvature and the slope of
the original potential, respectively.  The curvature is proportional to
$\tilde{k}^2$ and hence positive, irrespectively on the sign of $k$.
We also note that although $V(x)$ and $V'(x)$ are by construction
continuous, $V_\text{eff}(x)$ has a discontinuous jump in $x=b$. The
jump originates from the discontinuity of $V''(x)$ in $x=b$, which
gives $\Delta V_\text{eff} = V_\text{eff}(b^+) - V_\text{eff}(b^-) =
\widetilde{k}/2$. Although the potential in the Fokker-Planck problem is
repulsive, the effective potential of the associated quantum mechanical
problem is attractive and temperature dependent.

\begin{figure}[t]
\centering
\includegraphics[width=0.45\textwidth]{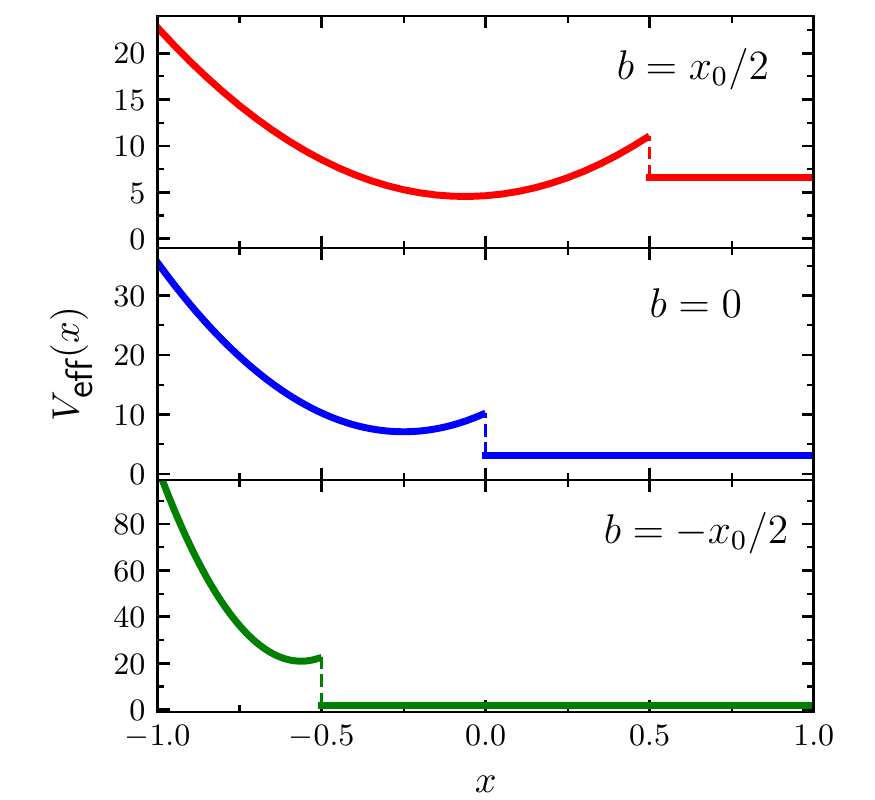}
\caption{Plot of $V_\text{eff} (x)$ for $x_0=1$, $\beta k=8$ and
three different values of $b$ ($b=0.5,0,-0.5$ from top to bottom).
We have here $\beta \Delta U = 4$, hence from Eq.~\eqref{bstar} we get
$b^* = x_0/3$, corresponding to the value of $b$ where the
minimum of the parabolic part of $V_\text{eff}$ is at the same
level as the constant part of the potential.}
\label{fig:Veff}
\end{figure}

Figure~\ref{fig:Veff} plots $V_\text{eff}(x)$ for three different values
of $b$ and fixed $k$, $\beta$ and $x_0$. For weak asymmetry ($b^* \leq
b \leq x_0$) the minimum of $V_\text{eff}$ is in the parabolic part of
$V(x)$. For $-x_0 < b \leq  b^*$ the minimum appears in the constant
part of $V_\text{eff}(x)$. We find
\begin{equation}
b^* = x_0 \, \frac{3- \sqrt{\beta \Delta U}}{1+\sqrt{\beta \Delta U}}
\label{bstar}
\end{equation}
Note that when the asymmetry is increased, $V_\text{eff}$ increases
in the parabolic region, while it decreases in the linear region.

Although $V_\text{eff}$ has a discontinuous jump, the solution $q(x)$ of
Eq.~\eqref{eq:FP2} and its derivative $q'(x)$ remain continuous in $x=b$.
To see this we integrate \eqref{eq:FP2} in an infinitesimal interval
$[b-\varepsilon,b+\varepsilon]$:
\begin{equation}
q'(b+\varepsilon) - q'(b-\varepsilon) = 
\int_{b-\varepsilon}^{b+\varepsilon} 
\left[ V_\text{eff}(x) - \lambda \right]\, q(x) \, dx
\end{equation}
As $\varepsilon \to 0$ the integral in the right hand side vanishes,
implying continuous derivative $q'(b^+) = q'(b^-)$ in $x=b$.
In conclusion we solve the eigenvalue problem requiring that $q(x)$
and $q'(x)$ to be continuous in $x=b$.

\section{The Eigenvalue Problem}

We solve \eqref{eq:FP2} separately in the two domains and then use 
the matching conditions of continuity of $q$ and $q'$ in $x=b$.

\subsection{The solution in the linear region ($b \leq x \leq x_0$)}

In this region $V_\text{eff}$ is constant, hence the solution 
vanishing at the boundary $q(x_0)=0$ is 
\begin{equation}
q(x) = B \sin \left[ \kappa \left( x_0 - x\right)\right]
\label{q_reg2}
\end{equation}
where $B$ is a constant and
\begin{equation}
\kappa^2 = \lambda - \frac{\widetilde{k}^2}{4} 
\left(b - \widetilde{x}\right)^2
\label{def:kappa}
\end{equation}
as obtained by plugging in \eqref{q_reg2} into \eqref{eq:FP2}.

\subsection{The solution in the parabolic region ($-x_0 \leq x \leq b$)}

As the parabolic potential is centered in $\widetilde{x}$ it is
convenient to shift the $x$-coordinate and define
\begin{equation}
Y(x) \equiv q( x + \widetilde{x} )
\end{equation}
Equation~\eqref{eq:FP2} then becomes
\begin{equation}
Y''(x) - \frac{\widetilde{k}^2 x^2}{4} \, Y(x) 
= -\left(\lambda - \frac{\widetilde{k}}{2}\right)\, Y(x)
\label{eq:FP3}
\end{equation}
The solutions of this equation are related to the Kummer confluent
hypergeometric functions $ _1F_1(a;b;z)$. The solutions even in $x$ are
\begin{equation}
Y_+(x) = e^{-\widetilde{k} x^2/4} \, _1F_1\left(\frac{s_n+1}{2};
\frac{1}{2};\frac{\widetilde{k} x^2}{2} \right)
\label{def:Y+}
\end{equation}
and the odd ones
\begin{equation}
Y_-(x) = \sqrt{\widetilde{k}} \, x\,e^{-\widetilde{k} x^2/4}   
\, _1F_1\left( \frac{s_n}{2}+1; \frac{3}{2};
\frac{\widetilde{k} x^2}{2}\right)
\label{def:Y-}
\end{equation}
We have defined here $s_n = -\lambda_n/\widetilde{k}$. The Kummer
hypergeometric functions are defined by
\begin{equation}
\, _1F_1 (a;b;z) = \sum_{k=0}^\infty  \frac{a^{(k)}}{b^{(k)} k!} \,  z^k
\label{app_def1F1}
\end{equation}
where $a^{(0)} = 1$, $a^{(k)} = a(a+1)(a+2)\ldots (a+k-1)$.  When the
argument $a$ is a negative integer, the sum in \eqref{app_def1F1}
becomes finite and reduces (apart some multiplicative factors) to
Hermite polynomials.

The most general solution is obtained by a linear combination of
\eqref{def:Y+} and \eqref{def:Y-}
\begin{equation}
q(x) = Y_+ (x - \widetilde{x}) + A Y_- (x - \widetilde{x}) 
\label{eq:linY}
\end{equation}
(without loss of generality, except in the symmetric limit $b=x_0$,
corresponding to $\widetilde{x}=0$, one can set one of the two
coefficients in \eqref{eq:linY} equal to $1$). Absorbing boundary
conditions impose a vanishing solution in $-x_0$ which yields
\begin{equation}
Y_+ (x_0 + \widetilde{x})  =  A Y_- (x_0 + \widetilde{x}) 
\label{zero}
\end{equation}
where we have used $Y_\pm (-x) = \pm Y_\pm (x)$. 

\subsection{Matching conditions}

Continuity of $q(x)$ and $q'(x)$ in $x=b$ implies
\begin{equation}
Y_{+} (b - \widetilde{x}) + A Y_{-} (b - \widetilde{x}) =  B \sin [ \kappa
(x_0 - b)] 
\label{funct}
\end{equation}
\begin{equation}
Y'_{+}  (b- \widetilde{x} ) + A Y'_{-}   (b-\widetilde{x})
 =  -\kappa B \cos [ \kappa (x_0 - b)]
\label{der}
\end{equation}
Combining \eqref{zero}, \eqref{funct} and \eqref{der} we can eliminate
$A$ and $B$ to obtain a single equation
\begin{equation}
\frac{Y_{-}(x_0+\widetilde{x}) Y_+(b-\widetilde{x}) + 
Y_+(x_0+\widetilde{x}) Y_-(b-\widetilde{x})}
{Y_{-}(x_0+\widetilde{x}) Y'_+(b-\widetilde{x}) + 
Y_+(x_0+\widetilde{x}) Y'_-(b-\widetilde{x})}
= - \dfrac{1}{\kappa} \tan [ \kappa (x_0 - b)] 
\label{numer}
\end{equation}
\eqref{numer} needs to be solved numerically to calculate the spectrum of
allowed values of $\lambda$, which is the only unknown in the equation.
Once $\lambda_n$ ($n=1,2 \ldots$) has been obtained one can use
\eqref{zero} to get $A$ and either \eqref{funct} or \eqref{der} to get
$B$. The eigenfunctions are then normalized to fulfill \eqref{eq:ortho}.

\begin{figure}[t!]
\begin{center}
\includegraphics[scale=1.0]{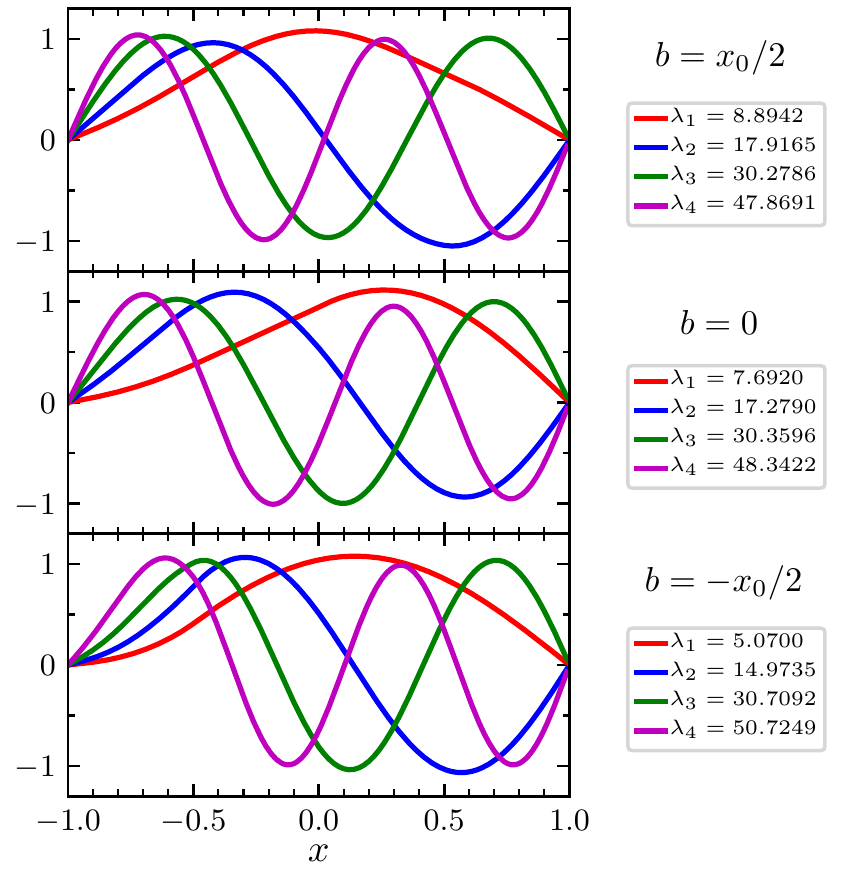}
\caption{Normalized eigenfunctions $q_n(x)$ for $n=1,2,3,4$ with $x_0=1$,
$\beta k=8$ (corresponding to $\beta \Delta U = 4$) and three different
values of $b$. The asymmetry has a stronger influence on the ground state
$q_1(x)$, while it has a milder effect on the higher excited states.}
\label{FIG:1F1}
\end{center}
\end{figure}

\subsection{Properties of the spectrum}

Figure~\ref{FIG:1F1} shows a list of the lowest eigenvalues $\lambda_n$
and the corresponding $q_n$ for fixed $k$, $x_0$, $\beta$ and for three
different values of $b$.  Note that the eigenfunctions have an increasing
number of nodes as the eigenvalue increases, as well-known in one
dimensional quantum mechanics. The nodes count provides a good check that
no eigenvalues are missed in the numerical calculation of the spectrum.
For sufficiently large $n$, quantum energies are high compared to the
variations of  $V_\text{eff}$ in the interval $[-x_0,x_0]$ and the wave
functions are close to that of a free particle in a box of width $2x_0$:
\begin{equation}
q_n(x) \approx \frac{1}{\sqrt{x_0}}  
\sin \left[ \pi n \frac{x+x_0}{2x_0}\right]
\label{eq:qn_diff}
\end{equation}
with eigenvalues $\lambda_n \approx \frac{\pi^2 n^2}{4 x_0^2}$.
The low-lying part of the spectrum is crucially dependent on the shape of
the potential, and this determines the long time behavior of the system.

\begin{figure}[t!]  
\begin{center}
\includegraphics[scale=1.0]{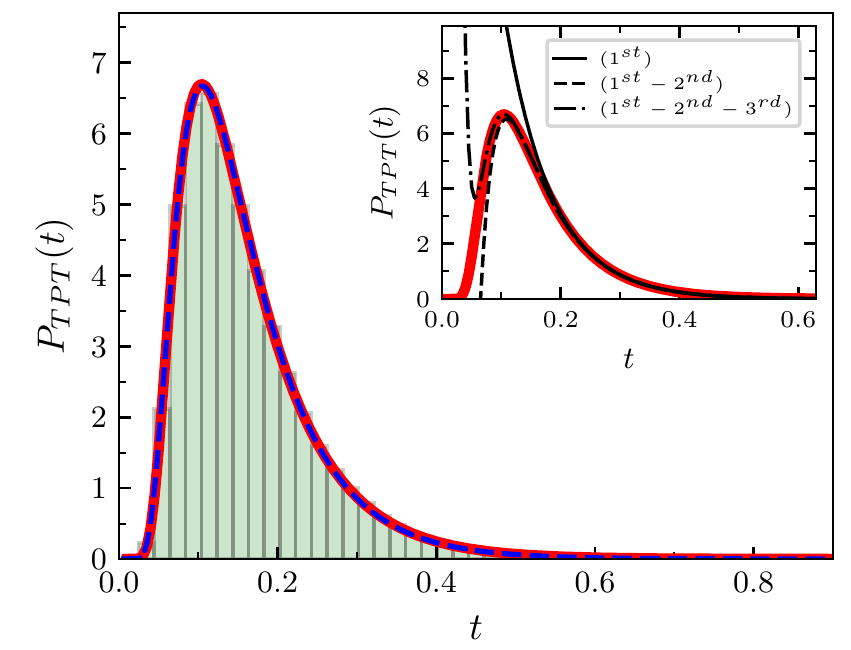} 
\caption{Red Solid line: TPT distribution for $x_0=1$, $\beta \Delta U =
4$, $b=-x_0/2$ and $D=2.5$ as obtained from Eq.\eqref{eq_TPTdist}. Green
bars: TPT histogram obtained from BD simulations. Blue Dashed line: TPT
histogram fitted from a parabolic barrier with $x_0=1$. The fit reproduces
very well the original data in the whole range, however with parameters
($\beta \Delta U = 1.8(1)$ and $D=2.81(1)$) which deviate significantly
from the original values.  Inset: Comparison between the full distribution
from Eq.\eqref{eq_TPTdist} (red solid line) and truncations of the same
expression to $n=1,2,3$ terms.}
\label{fig:test-TPT} 
\end{center} 
\end{figure}

\section{TPT distributions}

To obtain the TPT distribution for barrier crossing events from $-x_0$ to
$x_0$ one solves the Fokker-Planck equation \eqref{eq:FP} using a $\delta$
source in $-x_0+\varepsilon$. The probability current is given by
\begin{equation}
j_\varepsilon (x,t) = - D \left[ e^{-\beta V(x)} \partial_x
\left( e^{\beta V(x)} P_\varepsilon(x,t) \right) \right]
\end{equation}
with $P_\varepsilon(x,t)$ the solution with the said initial condition.
The TPT distribution is proportional to the current exiting from the
opposite boundary ($x_0$), hence
\begin{equation}
p_\text{TPT}(t) = \lim_{\varepsilon \to 0}
\frac{j_\varepsilon (x_0,t)}
{\int_0^\infty j_\varepsilon (x_0,t') \, dt'}
\end{equation}
This quantity can be written as follows (see e.g.\ Ref.\cite{cara18}
for a more detailed derivation):
\begin{equation} 
p_\text{TPT}(t) = -\frac{1}{A} 
\sum_{n=1}^{\infty} q_n'(-x_0) q_n'(x_0) e^{-\lambda_n D t}
\label{eq_TPTdist}
\end{equation}
where the constant 
\begin{equation}
A = - \sum_{n=1}^{\infty} q_n'(-x_0)q_n'(x_0)
\frac{1}{\lambda_n D}
\label{eq_normaliz}
\end{equation}
normalizes the distribution so that $\int_0^\infty p_\text{TPT}(t) \,
dt =1$. We note that the expression \eqref{eq_TPTdist} is symmetric with
respect to the direction of the paths, i.e. transitions $x_0 \to -x_0$
and $-x_0 \to x_0$ have the same distribution, reflecting a well-known
time reversal symmetry \cite{chau10} (holding also for asymmetric
barriers). Equation~\eqref{eq_TPTdist} is a series with coefficients
of alternating signs. This can be understood from the properties of
eigenfunctions. The ground state has no nodes hence the derivatives at
the two ends have opposite signs $q_1'(-x_0) q_1'(x_0) < 0$. The number
of nodes of $q_n(x)$ is equal to $n-1$, therefore in general $q_k'(-x_0)
q_k'(x_0) > 0$ for $k$ even and $q_k'(-x_0) q_k'(x_0) < 0$ for $k$ odd.

Figure~\ref{fig:test-TPT} shows a plot of a TPT distribution calculated
from Eq.~\eqref{eq_TPTdist} (red solid line) with $x_0=1$, $\beta
\Delta U =4$ and $D=2.5$. Data from Brownian Dynamics (BD) simulations
of barrier crossing events for the same set of parameters are shown as
green bars. The overlap between the two sets confirms the validity of
the numerical procedure to obtain the eigenvalues and eigenfunctions
from \eqref{numer}. While the BD simulations need large statistics, the
TPT distribution can be obtained more rapidly from \eqref{eq_TPTdist},
which is particularly advantageous in the high barrier limit. Moreover
from \eqref{eq_TPTdist} one can analyze separately the contribution
of each eigenstate to the total sum. In the calculations the sums
in \eqref{eq_TPTdist} and \eqref{eq_normaliz} are truncated to
$n=100$, a threshold value which guarantees very good accuracy of the
distribution. In Fig.~\ref{fig:test-TPT} we also plotted as a blue
dashed line the best fit to the data using a parabolic potential with
$x_0=1$, where $\beta k$ (or equivalently $\beta \Delta U$) and $D$
are taken as fitting parameters. The fitted curve overlaps extremely
well with the original data (showing that the TPT distribution for a
strongly asymmetric barrier is for practical purposes indistinguishable
from that generated by particles crossing a parabolic barrier), but
yields $\beta \Delta U = 1.8(1)$ and $D=2.81(1)$, which significantly
deviate from the original parameters used in the asymmetric potential. In
particular the barrier height is strongly underestimated, with respect
to the original value $\beta \Delta U =4$.  In order to infer the
contribution of the various terms in the sum \eqref{eq_TPTdist} to the
total TPT distribution we plotted in the inset of Fig.~\ref{fig:test-TPT}
again the full $p_\text{TPT}(t)$ from Eq.~\eqref{eq_TPTdist} (using the
first $100$ terms) and truncations to $n=1$, $2$ and $3$ terms. The
truncation is done only in the numerator of \eqref{eq_TPTdist},
while the normalization \eqref{eq_normaliz} is obtained from the full
calculation up to $n=100$. The analysis shows that the first three terms
in \eqref{eq_TPTdist} approximate well the whole decaying part of the
distribution and even the region beyond, but close to the maximum.

\begin{figure}[t!]
\begin{center}
\includegraphics[scale=1.0]{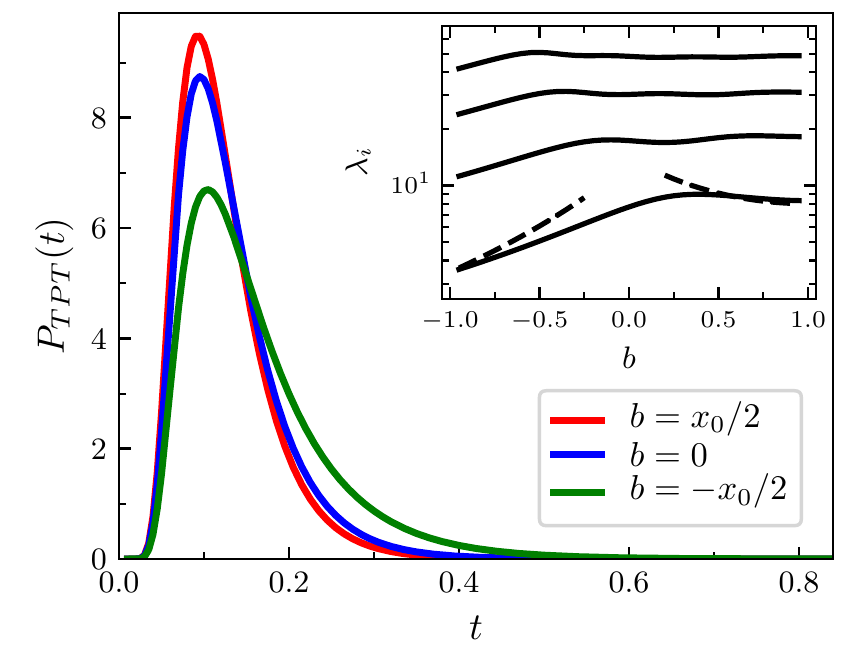}
\caption{Examples of TPT distributions $P_{TPT}(t)$ for $x_0=1$,
$\beta \Delta U=4$ ($\beta k=8$) and $D=2.5$. Average TPT increases
with increasing skewness of the energy barrier (i.e. decreasing
$b$). Typically, the distributions are obtained truncating the infinite
series \eqref{eq_TPTdist} to approx $n \leq 100$. Inset: behavior of
$\lambda_1$, $\lambda_2$, $\lambda_3$ and $\lambda_4$ (respectively
from bottom to top) as a function of $b$ for $x_0=1$ and $\beta \Delta
U=4$. The dashed line around $b \approx 1$ and $b \approx -1$ are
respectively $\lambda_1=\widetilde{k}$ and Eq.~\eqref{eq:l1_approx}
and approximate well the ground state energy around these two limits,
see text.}
\label{fig:TPT_exp}
\end{center}
\end{figure}

Figure~\ref{fig:TPT_exp} shows the TPT distribution as obtained from
Eq.~\eqref{eq_TPTdist} for the same parameters as in Fig.~\ref{fig:test-TPT}
and different values of $b$. In the strongest asymmetric case ($b=-x_0/2$)
the decay is the slowest and the average TPT is the highest of the three
cases analyzed. The inset of Fig.~\ref{fig:TPT_exp} plots the first four
eigenvalues for $-0.95 \leq b/x_0 \leq 0.95$, showing that the ground
state $\lambda_1$ is most strongly affected by variations of $b$, while
the influence of $b$ on higher eigenvalues is much weaker. For highly
excited eigenstates $\lambda_n$ converges to the free particle in a
box spectrum $\lambda_n \approx (\pi n/2x_0)^2$, which is independent
on $V_\text{eff}(x)$, and hence on $b$. The shape of the underlying
potential hence influences few low lying eigenstates. We also note that
$\lambda_1$ is a non-monotonic function of $b$, which is due to the two
competing trends on $V_\text{eff}(x)$ visible in Fig.~\ref{fig:Veff}. As
$b$ decreases from the symmetric case $b=x_0$ the parameter $\widetilde{k}$
defined in Eq.~\eqref{def:ktilde} increases.  For $b \approx x_0$ we can
approximate the ground state value to that of an harmonic oscillator in
an infinite domain $\lambda_1 = \widetilde{k}$. In the other limit of large
asymmetry $b \approx -x_0$, the ground state wavefunction is localized in
the constant part of $V_\text{eff}$ in the interval $b \leq x \leq x_0$
(see \eqref{def:Veff}). In first approximation
we can set the wavefunction to zero in $x=b$ and $x_0$, leading to
the ground state of a free quantum particle localized in $[b,x_0]$,
which is
\begin{equation}
\lambda_1 = \frac{\pi^2}{(x_0 - b )^2} + 
\frac{\widetilde{k}^2}{4} (b - \widetilde{x})^2
\label{eq:l1_approx}
\end{equation} 
Both $\lambda_1=\widetilde{k}$ and Eq.~\eqref{eq:l1_approx} approximate
well the ground state energy for $b \approx x_0$ and $b \approx -x_0$,
as shown by the dashed lines in the inset of Fig.~\ref{fig:TPT_exp},
and the transition between these two behaviors takes place around $b
\simeq b^*$ \eqref{bstar}, at which $\lambda_1$ is maximal.

\begin{table}
\caption{Summary of the analysis of TPT distributions generated from
Eq.~\eqref{eq_TPTdist} for $x_0=1$, $D=2.5$ and different values
of $\beta \Delta U$ and asymmetry parameter $b$. The second column gives
the average TPT obtained from the first moment of the distribution.
The last six columns report the parameters calculated from a fitting to a
$p_\text{TPT}(t)$ generated by a symmetric parabolic potential. The three
columns "ABC" refer to a fit with the absorbing boundary condition case,
that is Eq.~\eqref{eq_TPTdist} with $b=1$. As expected, the fit for $b=1$
reproduce the input values for $\beta \Delta U$, and $D$. The last three
columns are for a parabolic barrier with free boundary conditions given by
Eq.~\ref{eq_TPTdistNOAB}. The quality of the fit is estimated by $\chi$,
as defined by Eq.~\eqref{def_chi}. The overlapping distributions (solid
red line and the fitted dashed blue line) of Fig.~\ref{fig:test-TPT}
have a $\chi=10^{-3}$.  Uncertainties on $\langle t_\text{TPT} \rangle$,
and fitted $\beta \Delta U$, $D$ are approximately $\pm 1$ in the
last digit.}
\label{tab:SkewedPotential}
\begin{center}
\begin{small}
\begin{tabular}{c|c||ccc|ccc} 
\multicolumn{2}{c||}{$\beta \Delta U =4, D=2.5$} & \multicolumn{3}{c|}{fit to ABC} & 
\multicolumn{3}{c}{fit to FBC} \\
\hline
$b$&$\langle t_{TPT} \rangle$ &$\beta\Delta U$ & $D$ & $\chi$ & $\beta\Delta U$ & 
$D$ & $\chi$\\ 
\hline 
$1.0$  & $0.126$ & $4.0$ & $2.50$ & $0.000$ & $3.6$ & $3.11$ & $0.005$\\
$0.5$  & $0.123$ & $4.3$ & $2.44$ & $0.001$ & $3.9$ & $3.00$ & $0.003$\\
$0$    & $0.131$ & $3.6$ & $2.53$ & $0.001$ & $3.3$ & $3.15$ & $0.006$\\
$-0.5$ & $0.160$ & $1.8$ & $2.81$ & $0.001$ & $2.0$ & $3.66$ & $0.015$\\
\hline 
\hline
\multicolumn{2}{c||}{$\beta \Delta U =8, D=2.5$} & 
\multicolumn{3}{c|}{fit to ABC} & \multicolumn{3}{c}{fit to FBC} \\
\hline
$b$ & $\langle t_{TPT}\rangle$ &$\beta\Delta U$ & $D$ & $\chi$ & 
$\beta\Delta U$ & $D$ & $\chi$\\ 
\hline 
$1.0$  & $0.082$ & $8.0$ & $2.50$ & $0.000$ & $7.1$ & $2.90$ & $0.003$\\
$0.5$  & $0.079$ & $9.1$ & $2.35$ & $0.004$ & $8.1$ & $2.69$ & $0.001$\\
$0$    & $0.083$ & $8.4$ & $2.38$ & $0.006$ & $7.4$ & $2.76$ & $0.002$\\
$-0.5$ & $0.104$ & $4.5$ & $2.82$ & $0.001$ & $4.0$ & $3.35$ & $0.010$\\
\hline
\hline
\end{tabular}
\end{small}
\end{center}
\end{table}

Table~\ref{tab:SkewedPotential} summarizes the results of the analysis.
TPT distributions were generated from Eq.~\eqref{eq_TPTdist} with
parameters $x_0=1$, $D=2.5$, $\beta \Delta U=4$ and $\beta \Delta
U=8$ and four different values of $b$. These data were fitted to TPT
distributions generated from \eqref{eq_TPTdist} but with a parabolic
potential, e.g. $b=x_0=1$. Another fit was performed against the
distribution~\cite{zhan07,lale17}
\begin{equation} 
p_\text{TPT}(t)= -\dfrac{2}{\pi} \dfrac{G\,'(t) 
\, e^{-G^2(t)}}{1-\text{Erf}[\sqrt{\beta \Delta U}]} \; ,
\label{eq_TPTdistNOAB}
\end{equation}
where
\begin{equation} 
G(t) = \sqrt{\beta \Delta U} 
\sqrt{\frac{e^{\Omega t}+1}{e^{\Omega t}-1}}
\label{eq_defG}
\end{equation}
and where $\Omega = \beta k D$ is the characteristic rate of the process.
Equation~\eqref{eq_TPTdistNOAB} is the TPT distribution for a particle
crossing a parabolic potential, obtained from the free boundary condition case.
This simpler analytical expression (compared to the infinite series
\eqref{eq_TPTdist}) is most commonly used to fit experimental data
\cite{neup16}. For very high barrier multiple crossing at boundaries are
highly unlikely and the free boundary conditions distribution provides
a very good approximation to the exact one.  As check of the quality of
the fits the following parameter was used
\begin{equation}
\chi = \int_0^{\infty} \left[p_\text{TPT}(t)-
p_\text{TPT}^\text{(fit)}(t)\right]^2 p_\text{TPT}(t) \, dt
\label{def_chi}
\end{equation}
In all cases analyzed, and as shown in Fig.~\ref{fig:test-TPT} the TPT
distributions are well-fitted by distributions obtained from a parabolic
potential, as demonstrated by the very low value of $\chi$ reported in
Table~\ref{tab:SkewedPotential}. However, as discussed in the case of
Fig.~\ref{fig:test-TPT}, the fitted values can significantly deviate
from the original ones. Table~\ref{tab:SkewedPotential} shows that the
effect is most significant for estimates of the barrier height.

\section{Discussion}

Transition paths are events of very short duration (typically in the $\mu
s$ scale for protein and nucleic acid folding \cite{chun12}), therefore
their experimental study has been very challenging for long time. Recent
advances in single molecule techniques, have allowed the determination
of average TPT in biomolecular folding experiments \cite{chun12,neup12}
and in some cases also the full TPT distribution \cite{neup16}. The
comparison of experiments with theory has revealed that some quantities
match quite well the simple barrier crossing model. In particular,
estimates of the diffusion coefficient, as obtained from the analysis
of average TPT, their distribution, shapes and velocities, turned out to
be quite robust and consistent among each other \cite{neup18}. However,
a markable disagreement was found in the determination of the barrier
height \cite{neup16}, where a fit of $p_\text{TPT}(t)$ (as given by
Eq.~\eqref{eq_TPTdistNOAB}) gave a very small barrier height $\beta
\Delta U \approx 0.5$, as compared to other thermodynamic measurements,
from which $\beta \Delta U \approx 9$ was estimated. This large difference
has triggered several theoretical studies to reconcile the measurements
with theory \cite{maka17,poll16,sati17}. Motivated by the discrepancy
we investigated in this paper TPT distribution for asymmetric potential
barriers composed by a parabolic part, joined to a linear part. Using
the mapping of the Fokker-Planck equation to a one-dimensional quantum
problem, we expressed the TPT distribution as an eigenfunctions expansion.

Our analysis showed that TPT distributions generated from asymmetric
potentials are for all practical purposes indistinguishable from those
obtained from a parabolic potential. Fitting the former against the latter
returns two values as fitting parameters: the scaled barrier height $\beta
\Delta U$ and the diffusion constant $D$. The results, summarized in
Table~\ref{tab:SkewedPotential}, show that these parameters deviate from
the input values. For instance, for an asymmetric barrier with $b=-0.5$,
$\beta \Delta U =4$, $D=2.5$, the fit to a TPT distribution with a
parabolic potential and absorbing boundaries, gives $\beta \Delta U =1.8$
and $D=2.8$. While the variation in $D$ is of about $10\%$, there is more
than a factor $2$ of difference in the estimated barrier height, which is
somehow reminiscent to what one observes in experiments~\cite{hoff19} when
fitting empirical distributions with Eq.~\eqref{eq_TPTdistNOAB} (although
the reported difference is much higher in that case \cite{neup16}). Our
work suggests that, while $p_\text{TPT}(t)$ is somehow universal,
to get a reliable estimate of $\beta \Delta U$ and $D$ from a given
empirical distribution one needs to know the shape of the underlying
potential. This is particularly true when estimating $\beta \Delta U$
and can be understood as follows. At long times, $p_\text{TPT}(t)$ is
determined by few low-lying eigenvalues of the associated Schr\"odinger
equation (see inset of Fig.~\ref{fig:test-TPT}). These eigenvalues are
most strongly affected by the shape of $V_\text{eff}(x)$. To understand
this let us consider the average TPT for a parabolic potential and
free boundaries
\begin{equation} 
\langle t_\text{TP} \rangle = \tau_\text{diff} \frac{\log
(\beta \Delta U)}{\beta \Delta U} 
\label{aveTPT_parab} 
\end{equation}
where $\tau_\text{diff} = 4 x_0^2/D$ is the characteristic diffusion time
for a free particle to cross the interval $[-x_0,x_0]$. This relation,
valid for $\beta \Delta U \gg 1$ and originally derived by A. Szabo
\cite{chun12}, can also be obtained as first moment of the distribution
\eqref{eq_TPTdistNOAB}, see e.g. Ref.~\cite{lale17}. Assuming that the
average TPT is fully determined by the lowest eigenvalue so that $\langle
t_\text{TP} \rangle \sim \lambda_1^{-1}$, Eq.~\eqref{aveTPT_parab} then
implies that for a parabolic potential $\lambda_1 \sim \beta \Delta U$.
As we have shown here, the shape of the potential landscape $V(x)$ and
as a consequence of $V_\text{eff}(x)$, has a rather strong influence on
$\lambda_1$ (see Fig.~\ref{FIG:1F1} and inset of Fig.~\ref{fig:TPT_exp}),
it is therefore not surprising that fits of empirical TPT may lead to
rather strong deviations of $\beta \Delta U$.

The diffusion constant is instead much less influenced by the
shape of the potential. Although the average TPT depends on $D$,
see Eq.~\eqref{aveTPT_parab}, its value in fitting empirical data is
mostly determined by the short time behavior of the TPT distribution
\cite{zhan07,lale17}, which is in turn dependent on the spectrum of
eigenvalues $\lambda_n$ for large $n$. But these eigenvalues, as discussed
before, are only very weakly affected by the shape of $V(x)$, and
essentially fixed by the vanishing wavefunction at the absorbing
boundaries, thus  converge to the free particle in a box spectrum.

In conclusion our work shows that a reliable estimate of the barrier
height from the analysis of $p_\text{TPT}(t)$ can only be obtained if one
has some knowledge of the shape of the potential. Note that this is less
problematic when estimating $\beta \Delta U$ from reaction rates. Kramers'
formula for barrier crossing rates \cite{hang90} shows that the shape
of the potential only weakly affect the total rate compared to the
exponential dependence on $\beta \Delta U$. Recent experiments report
a $\beta \Delta U = 9$ from Kramers theory and $\beta \Delta U \approx
0.5$ from a fit of TPT distribution. This discrepancy is larger than
those reported in Table~\ref{tab:SkewedPotential}. It is likely that
other effects also play a role in giving rise to this large difference,
see \cite{poll16,sati17,maka17}, but certainly barrier asymmetry should
be taken into account in coming studies.

\section*{Conflicts of interest}
There are no conflicts to declare.

\section*{Acknowledgements}

We acknowledge interesting and stimulating discussions with Prof. Carlo
Vanderzande, who passed away on September 2, 2019. MC acknowledges
financial support from KU Leuven grant C12/17/006 and from the Austrian
Science Fund (FWF): P 28687-N27. TS acknowledges financial support from
JSPS KAKENHI (Grant No. JP18H05529) from MEXT, Japan, and JST, PRESTO
(Grant No. JPMJPR16N5). EC is grateful to the Department of Physics and
Mathematics of the Aoyama Gakuin University, where part of this work
was done, for kind hospitality.

\balance



\providecommand*{\mcitethebibliography}{\thebibliography}
\csname @ifundefined\endcsname{endmcitethebibliography}
{\let\endmcitethebibliography\endthebibliography}{}

\end{document}